\documentclass[aps,prl,twocolumn,superscriptaddress]{revtex4-1}%
\usepackage{amsmath,amssymb,amsbsy,subfigure,hyperref,subfigure,bbm,times,txfonts,accents,float}
\usepackage{graphicx}
\usepackage{dcolumn}
\usepackage{bm}

\begin{document}

\title{A Unified and Strengthened Framework for the Uncertainty Relation}

\author{Xiao Zheng}
\affiliation{%
 Key Laboratory of Micro-Nano Measurement-Manipulation and Physics (Ministry of Education), School of Physics and Nuclear Energy Engineering, Beihang University, Xueyuan Road No. 37, Beijing 100191, China
}%
\author{Shao-Qiang Ma}
\affiliation{%
 Key Laboratory of Micro-Nano Measurement-Manipulation and Physics (Ministry of Education), School of Physics and Nuclear Energy Engineering, Beihang University, Xueyuan Road No. 37, Beijing 100191, China
}%
\author{Guo-Feng Zhang}%
 \email{gf1978zhang@buaa.edu.cn}
\affiliation{%
 Key Laboratory of Micro-Nano Measurement-Manipulation and Physics (Ministry of Education), School of Physics and Nuclear Energy Engineering, Beihang University, Xueyuan Road No. 37, Beijing 100191, China
}%

\author{Heng Fan}

\affiliation{
 Beijing National Laboratory for Condensed Matter Physics, and Institute of Physics,
Chinese Academy of Sciences, Beijing 100190, China
}%
\author{Wu-Ming Liu}

\affiliation{
 Beijing National Laboratory for Condensed Matter Physics, and Institute of Physics,
Chinese Academy of Sciences, Beijing 100190, China
}%

\date{\today}

\begin{abstract}
We provide a unified and strengthened framework for the product form and the sum form variance-based uncertainty relations by constructing a unified uncertainty relation. In the unified framework, we deduce that the uncertainties of the incompatible observables are bounded by not only the commutator of themselves, but also the quantities related with the other operator. This operator can provide information so that we can capture the uncertainty of the measurement result more accurately, and thus is named as the information operator. The introduction of the information operator can fix the deficiencies in both the product form and the sum form uncertainty relations, and provides a more accurate description of the quantum uncertainty relation. The unified framework also proposes a new interpretation of the uncertainty relation for non-Hermitian operators; i.e., the ``observable" second-order origin moments of the non-Hermitian operators cannot be arbitrarily small at the same time when they are generalized-incompatible on the new definition of the generalized commutator.
\end{abstract}
\maketitle
 Quantum uncertainty relations \cite{1,2,3}, expressing the impossibility of the joint sharp preparation of the incompatible observables \cite{4w,4}, are the most fundamental differences between quantum  and classical mechanics \cite{5,6,7,28S}. The uncertainty relation has been widely used in the quantum information science, such as quantum non-cloning theorem \cite{7P,7H}, quantum cryptography \cite{CW,C7,8,9}, entanglement detection \cite{10,10C,24,11,12}, quantum spins squeezing \cite{13,14,15,16}, quantum metrology \cite{7V,17,18}, quantum synchronization \cite{18A,18F} and mixedness detection \cite{19,20}. In general, the improvement in uncertainty relations will greatly promote the development of quantum information science \cite{7V,10,20B,21P,21X}.

The variance-based uncertainty relations for two incompatible observables $A$ and $B$  can be divided into two forms: the product form ${\Delta A}^2{\Delta B}^2\geq LB_{p}$ \cite{2,3,4,28L} and the sum form ${\Delta A}^2+{\Delta B}^2\geq LB_{s}$ \cite{21,22,23,25}, where $LB_{p}$ and $LB_s$ represent the lower bounds of the two forms uncertainty relations, and ${\Delta Q}^2$ is the variance of $Q$ \cite{25F}. The product form uncertainty relation cannot fully capture the concept of incompatible observables, because it can be trivial; i.e., the lower bound $LB_{p}$ can be null even for incompatible observables \cite{21,22,4JL,28}. This deficiency is referred to as the triviality problem of the product form uncertainty relation. In order to fix the triviality problem,  Maccone and Pati deduced a sum form uncertainty relation with a nonzero lower bound for incompatible observables \cite{28}, showing that the triviality problem can be addressed by the sum form uncertainty relation.  Since then, lots of effort has been made to investigate the uncertainty relation in the sum form  \cite{10,21,26,27,H2,42}.  However, most of the sum form uncertainty relations depend on the orthogonal state to the state of the system, and thus are difficult to apply to the high dimension Hilbert space \cite{21}. There also exist the uncertainty relations based on the entropy \cite{5,6,7P,34} and skew information \cite{34S}, which may not suffer the triviality problem, but they cannot capture the incompatibility in terms of the experimentally measured error bars, namely variances \cite{28,28S}.

Here we only focus on the uncertainty relation based on the variance. Despite the significant progress on the variance-based uncertainty relation, previous work mainly studies the product form and the sum form uncertainty relations, separately. A natural question is raised : can the uncertainty relations in the two forms be integrated into a unified framework? If so, can the unified framework fix the deficiencies in the traditional uncertainty relations and provide a more accurate description of the quantum uncertainty relation? In other words, can the unified framework provide a stronger theoretical system for the quantum uncertainty relation?

In this Letter, we provide a unified framework for the product form and the sum form variance-based uncertainty relations by constructing a unified uncertainty relation. The unified framework shows that the uncertainties of the incompatible observables $A$ and $B$ are bounded by not only their commutator, but also the quantities related with the other operator, named as the information operator. Actually, the deficiencies in both the product form and the sum form uncertainty relations can be considered as having not taken the information operator into consideration and can be completely fixed by the introduction of the information operator. Furthermore, the uncertainty inequality will become an uncertainty equality when a specific number of information operators are introduced, which means the uncertainty relation can be expressed exactly with the help of the information operators. Thus the unified framework provides a strengthened theoretical system for the uncertainty relation. Meanwhile, our uncertainty relation provides a new interpretation of the uncertainty relation for non-Hermitian operators; i.e., the ``observable" second-order origin moments of the non-Hermitian operators cannot be arbitrarily small at the same time when they are generalized-incompatible on the new definition of the generalized commutator. The new interpretation reveals some novel quantum properties that the traditional uncertainty relation cannot do.

\emph{Unified Uncertainty Relation.--- }The Schr\"{o}dinger uncertainty relation (SUR) is the initial as well as the most widely used product form uncertainty relation \cite{3}:
\begin{align}
{\Delta A}^2{\Delta B}^2\geq\frac{1}{4}|\langle[A,B]\rangle|^2+\frac{1}{4}|\langle\{\check{A},\check{B}\}\rangle|^2\tag{1},
\end{align}
where $\langle Q\rangle$ represents the expected value of $Q$,  $\check{Q}=Q-\langle Q\rangle$,  $[A,B]=AB-BA$ and $\{\check{A},\check{B}\}=\check{A}\check{B}+\check{B}\check{A}$ represent the commutator and anti-commutator, respectively. One of the most famous sum form uncertainty relations, which have fixed the triviality problem in the product form uncertainty relation, takes the form \cite{28}:
\begin{align}
{\Delta A}^2+{\Delta B}^2\geq|\langle\psi|A\pm iB|\psi^\perp\rangle|^2\pm i\langle[A,B]\rangle \tag{2},
\end{align}
where $|\psi^\bot\rangle$ is the state orthogonal to the state of the system $|\psi\rangle$.

Before constructing the unified uncertainty relation, we first consider the non-Hermitian extension of the commutator and anti-commutator. There exist two kinds of operators in quantum mechanics: Hermitian and non-Hermitian operators, but it should be paid particular attention that lots of uncertainty relations are invalid for non-Hermitian operators \cite{37,38,39}. For instance,  $|[\sigma_+,\sigma_-]|^2/4+|\{\check{\sigma}_+,\check{\sigma}_-\}|^2/4\geq{\Delta\sigma_+}^2{\Delta\sigma_-}^2$ , where the non-Hermitian operator $\sigma_+(\sigma_-)$ is the raising (lowering) operator of the single qubit system. That is to say,  different from the Hermitian operators, the uncertainties of the non-Hermitian operators are not lower-bounded by quantities related with the commutator. The essential reason for this phenomenon is that  $i[\mathcal{A},\mathcal{B}]$ and $\{\mathcal{A},\mathcal{B}\}$ cannot be guaranteed to be Hermitian by the existing definition of commutator and anti-commutator when the operator $\mathcal{A}$ or $\mathcal{B}$ is non-Hermitian. To fix this problem, we define the generalized commutator and anti-commutator as:
\begin{align}
[\mathcal{A},\mathcal{B}]_{\mathcal{G}}=\mathcal{A}^\dag\mathcal{B}-\mathcal{B}^\dag\mathcal{A},  \quad \{\mathcal{A},\mathcal{B}\}_{\mathcal{G}}=\mathcal{A}^\dag\mathcal{B}+\mathcal{B}^\dag\mathcal{A}\tag{3}.
\end{align}
The generalized commutator and anti-commutator will reduce to the normal ones when $\mathcal{A}$ and $\mathcal{B}$ are both Hermitian. We say that $\mathcal{A}$  and $\mathcal{B}$ are generalized-incompatible (generalized-anti-incompatible) with each other hereafter when $\langle[\mathcal{A},\mathcal{B}]_{\mathcal{G}}\rangle\neq0$  $(\langle\{\mathcal{A},\mathcal{B}\}_{\mathcal{G}}\rangle\neq0)$. Then, one can obtain a new uncertainty relation for both Hermitian and non-Hermitian operators (for more detail, please see the Unified Uncertainty Relation in the Supplemental Material \cite{35}):
\begin{align}
\langle\mathcal{A}^\dag\mathcal{A}\rangle\langle\mathcal{B}^\dag\mathcal{B}\rangle=\frac{|\langle [{\mathcal{A}},{\mathcal{B}}]_{\mathcal{G}}\rangle|^2}{4}+\frac{|\langle\{{\mathcal{A}},{\mathcal{B}}\}_{\mathcal{G}}\rangle|^2}{4}+\langle\mathcal{C}^\dag\mathcal{C}\rangle\langle\mathcal{B}^\dag\mathcal{B}\rangle \tag{4},
\end{align}
where the remainder $\langle\mathcal{C}^\dag\mathcal{C}\rangle\langle\mathcal{B}^\dag\mathcal{B}\rangle\geq0$ with $\mathcal{C}=\mathcal{A}-\langle\mathcal{B}^\dag\mathcal{A}\rangle\mathcal{B}/\langle\mathcal{B}^\dag\mathcal{B}\rangle$, and $\langle\mathcal{Q}^\dag\mathcal{Q}\rangle$ is the second-order origin moment of the operator $\mathcal{Q}$.

In fact, the traditional interpretation of the uncertainty relation is invalid for non-Hermitian operators, because, as mentioned above,  most of the uncertainty relations will be violated when applied to non-Hermitian operators. The uncertainty relation (4) provides a new interpretation of the uncertainty relation for non-Hermitian operators; i.e., the second-order origin moments $\langle\mathcal{A}^\dag\mathcal{A}\rangle$ and $\langle\mathcal{B}^\dag\mathcal{B}\rangle$ cannot be arbitrarily small at the same time when $\mathcal{A}$ and $\mathcal{B}$ are generalized-incompatible or generalized-anti-incompatible with each other. Remarkably, the operators $\mathcal{A}^\dag\mathcal{A}$, $\mathcal{B}^\dag\mathcal{B}$, $i[\mathcal{A},\mathcal{B}]_{\mathcal{G}}$, and $\{\mathcal{A},\mathcal{B}\}_{\mathcal{G}}$ are Hermitian even when $\mathcal{A}$ and $\mathcal{B}$ are non-Hermitian. That is to say, different from the variance, the second-order origin moment is observable for both the Hermitian and non-Hermitian operators. The new interpretation reveals some novel quantum properties that the traditional uncertainty relations cannot do. Such as, applying the new uncertainty relation (4) to the annihilation operators $a_1$ and $a_2$ of two continuous variable subsystems, one can deduce that the product of the expected energy of two subsystems $\langle a_1^\dag a_1\rangle\langle a_2^\dag a_2\rangle$ is lower-bounded by   $|\langle[a_1,a_2]_{\mathcal{G}}\rangle|^2/4+|\langle\{a_1,a_2\}_{\mathcal{G}}\rangle|^2/4$. Especially, the energy of two subsystems cannot be arbitrarily small at the same time, when the annihilation operators of the two systems are generalized-incompatible or generalized-anti-incompatible on the state of the system, which means $\langle[a_1,a_2]_{\mathcal{G}}\rangle$ or $\langle\{a_1,a_2\}_{\mathcal{G}}\rangle$ does not equal or tend to zero.

  The new uncertainty relation (4) expresses the quantum uncertainty relation in terms of the second-order origin moment, instead of the variance, but can unify the uncertainty relations based on the variance.  Then, we demonstrate that some well-known uncertainty relations in either the sum form or the product form can be unified by the new uncertainty relation. Firstly, the new uncertainty relation turns into the product form uncertainty relation SUR, if we replace the operators $\mathcal{A}$ and $\mathcal{B}$ with the Hermitian operators $\check{A}=A-\langle A\rangle$ and $\check{B}=B-\langle B\rangle$. Secondly, assuming the system is in the pure state $|\psi\rangle$ and substituting the non-Hermitian operators  $\mathcal{A}=\check{A}\pm i\check{B}$ and $\mathcal{B}=|\psi^\bot\rangle\langle\psi|$ into the uncertainty relation  (4), one can obtain the sum form uncertainty relation (2). Here, the product form $\langle\mathcal{A}^\dag\mathcal{A}\rangle\langle\mathcal{B}^\dag\mathcal{B}\rangle=\Delta(A\pm iB)^{2}\Delta(|\psi^\bot\rangle\langle\psi|)^{2}=\Delta A^{2}+\Delta B^{2}\pm i\langle[A,B]\rangle$ turns into the sum form. That is to say, the product form uncertainty relation is the new uncertainty relation for Hermitian operators and the sum form uncertainty relation is actually the new uncertainty relation for non-Hermitian operators. The other uncertainty relations in the two forms \cite{26,42,27,41,37,38,23,28} can also be recovered by the uncertainty relation (4) in the similar way, and thus the equality (4) provides a unified uncertainty relation.

   The uncertainty relations in the two forms can be divided into several categories with respect to their purposes and applications, such as the uncertainty relations focused on the effect of the incompatibility of observables on the uncertainty \cite{26,42}, the uncertainty relations used to investigate the relation between the variance of the sum of the observables and the sum variances of the observables \cite{23,28}, and even the uncertainty relations for three and more observables \cite{27,41}. The unified uncertainty indicates that they can all be integrated into a unified framework. Besides,  by the introduction of the information operator, the unified framework provides a strengthened theoretical system for the quantum uncertainty relation. That is to say, the unified framework can fix the deficiencies in the traditional uncertainty relations, and provides a more accurate description of the uncertainty relation. The corresponding discussion will be presented in the next section.

\begin{figure}
\centering 
\includegraphics[height=4.5cm]{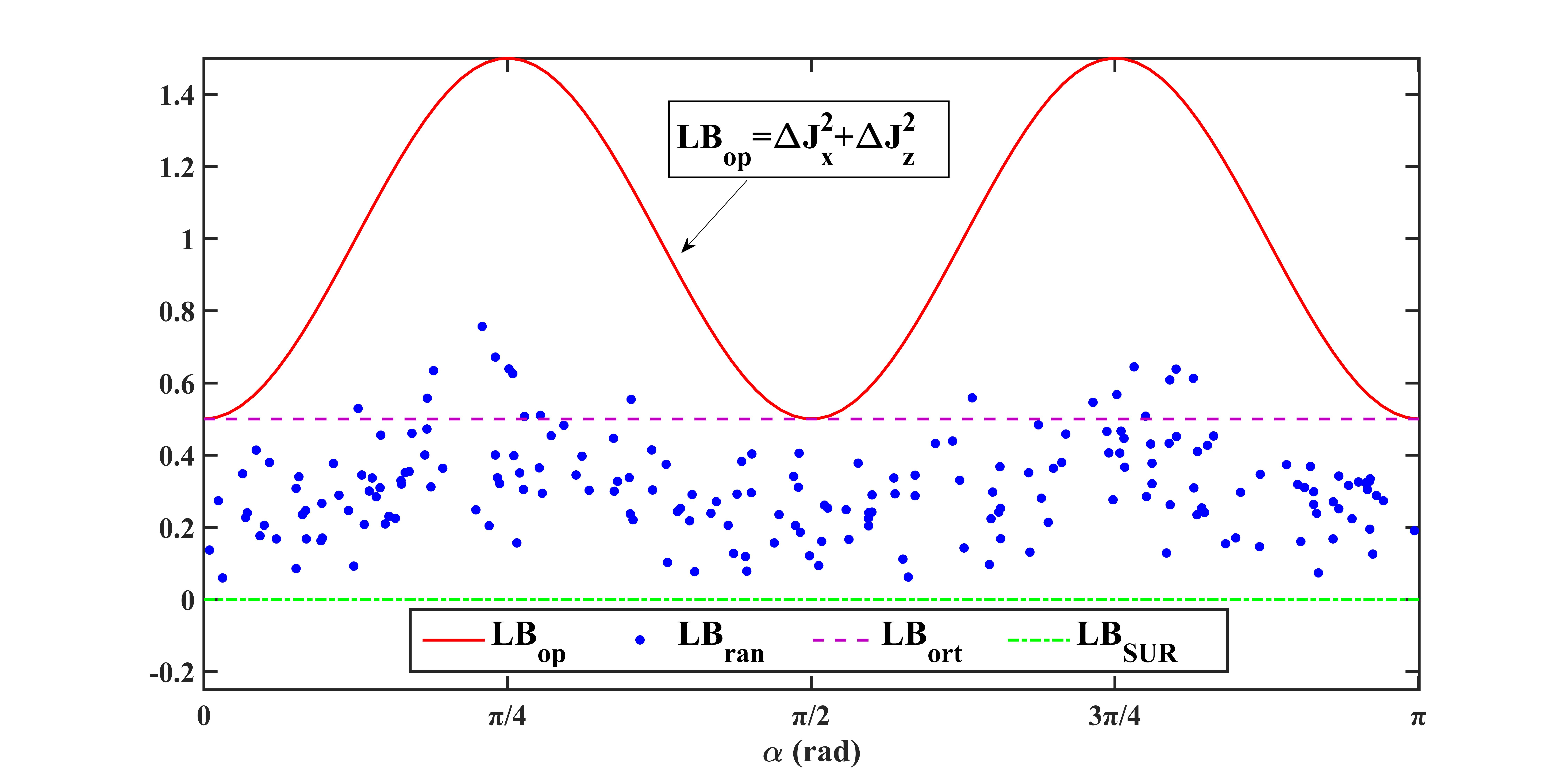}
\caption{The spin-1 system is chosen as the platform to demonstrate the new uncertainty inequality (8). We take $A=J_x$, $B=J_z$, $\hbar=1$ , and the state is parameterized by $\alpha$ as $\rho=\cos^2(\alpha)|1\rangle\langle1|+\sin^2(\alpha)|-1\rangle\langle-1|$, with $|\pm1\rangle$ and $|0\rangle$ being the eigenstates of $J_z$. The green dash-dotted line represents the lower bound of the SUR (denoted by $LB_{SUR}$). It can be seen that the lower bound of the SUR is trivial all the time. According to Ref.\cite{28}, the uncertainty (2) turns into ${\Delta A}^2+{\Delta B}^2\geq \mp i\langle[A,B]\rangle+\langle(-A\pm iB|\psi^\bot\rangle\langle\psi^\bot|(-A\mp iB)\rangle$ for the mixed state $\rho_{mixed}=\sum p_j|\psi_j\rangle\langle\psi_j|$, if there exists a state $|\psi^\bot\rangle$ orthogonal to all states $|\psi_j\rangle$. Obviously, the orthogonal state $|\psi^\bot\rangle$ can only be taken as $|0\rangle$ for the given state $\rho$, and the corresponding bound is noted by the purple dashed line  (denoted by $LB_{ort}$). The 200 blue dots (denoted by $LB_{ran}$) stands for the lower bound (8) which are calculated by randomly taking 200 sets of  $\alpha$ , $\mathcal{R}$ and $\mathcal{S}$ into (8). The red solid line is the optimal lower bound of (8) (denoted by $LB_{op}$),  which is obtained by taking $\mathcal{R}$ or $\mathcal{S}=\lambda_1\check{A}+\lambda_2\check{B}$ with $|\lambda_1|^2=|\lambda_2|^2\neq0$. We can find that $LB_{op}$ is exactly equal to the sum of the uncertainty $\Delta J_x^2+\Delta J_z^2$. } 
\end{figure}

\begin{figure}
\centering 
\includegraphics[height=4.5cm]{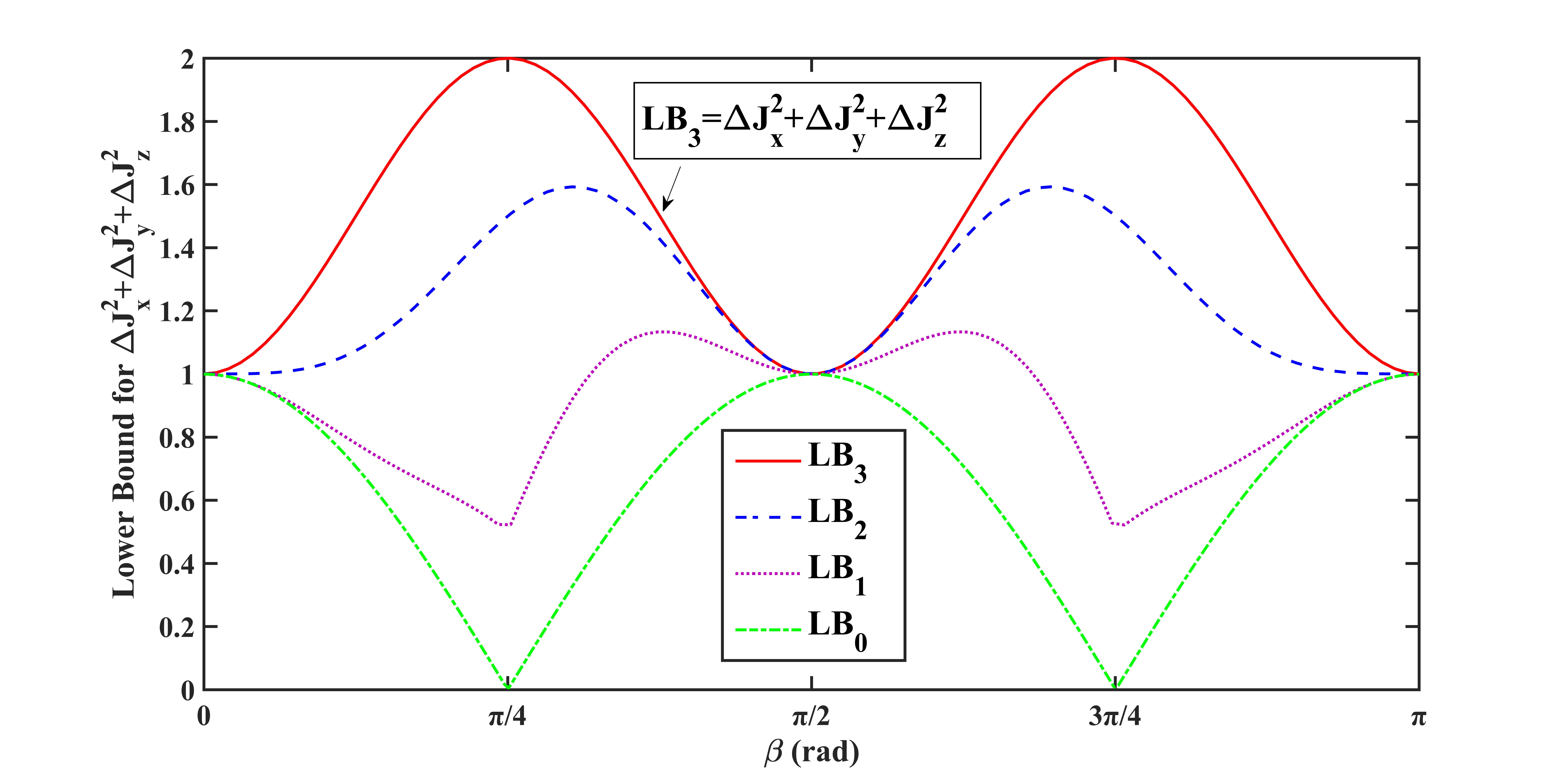}
\caption{Illustration to demonstrate the function of the information operator is presented. We take $\hbar=1$, and assume that the state of the spin-1 system is in the pure state $|\psi\rangle=\cos(\beta)|1\rangle+\sin(\beta)|-1\rangle$ with $|\pm1\rangle$ being the eigenstates of $J_z$. The operator set $\Theta=\{\mathcal{O}_1,\mathcal{O}_2, \mathcal{O}_3\}$ can be obtained by the Schmidt transformation ( see the Schmidt transformation process in the Supplemental Material \cite{35}). The uncertainty relation (9) turns into a series of sum form uncertainty relations: $\Delta J_x^2+\Delta J_y^2+\Delta J_z^2\geq LB_K=\sum_{k=1}^{K}|\langle\mathcal{O}_k^\dag(e^{i\theta_1}J_x+e^{i\theta_2}J_y+e^{i\theta_3}J_z)\rangle|^2/\langle\mathcal{O}_k^\dag\mathcal{O}_k\rangle-\langle\{e^{i\theta_1}J_x,e^{i\theta_2}J_y\}_{\mathcal{G}}\rangle-\langle\{e^{i\theta_2}J_y,e^{i\theta_3}J_z\}_{\mathcal{G}}\rangle-\langle\{e^{i\theta_3}J_z,e^{i\theta_1}J_x\}_{\mathcal{G}}\rangle$ with $K=1,2,3$, when we take $A_1=J_x, A_2=J_y, A_3=J_z$ and $X=\{e^{i\theta_1}, e^{i\theta_2}, e^{i\theta_3}\}^T$. Obviously, the tightness of the uncertainty relation becomes better and better with $K$ increasing, and the uncertainty inequality will become an equality when $K=3$. The uncertainty relation becomes $\Delta J_x^2+\Delta J_y^2+\Delta J_z^2\geq LB_0=-\langle\{e^{i\theta_1}J_x,e^{i\theta_2}J_y\}_{\mathcal{G}}\rangle-\langle\{e^{i\theta_2}J_y,e^{i\theta_3}J_z\}_{\mathcal{G}}\rangle-\langle\{e^{i\theta_3}J_z,e^{i\theta_1}J_x\}_{\mathcal{G}}\rangle$  when we do not take the information operators into consideration. It is clear that the tightness of $LB_0$ is worse than the other uncertainty relations, and $LB_0$ is trivial when $\beta=\pi/4$ and $3\pi/4$.
} 
\end{figure}

\emph{Information operator.---}Based on the initial spirit of Schr\"{o}dinger, the SUR can be derived as follows \cite{25V}. Assume $\mathcal{F}=\sum^N_{m=1}x_m\check{A}_m$, where $A_m$ stands for an arbitrary operator, $N$ is the number of the operators and $x_m\in C$ represents a random complex number. Using the non-negativity of the second-order origin moment of $\mathcal{F}$ \cite{25V}, namely $\langle \mathcal{F}^\dag \mathcal{F}\rangle\geq0$, one can obtain:
\begin{align}
&\mathbb{D}:\geq0\tag{5},
\end{align}
where $\mathbb{D}$ is the $N\times N$ dimension matrix with the elements $\mathbb{D}(m,n)=\langle\check{A}_m^\dag\check{A}_n\rangle$ and $\mathbb{D}:\geq0$ means that  $\mathbb{D}$ is a positive semidefinite matrix.  As for the positive semidefinite matrix $\mathbb{D}$, we have ${\rm Det}(\mathbb{D})\geq0$ with ${\rm Det}(\mathbb{D})$ being the determinant value of ${\mathbb{D}}$, and $X^\dag.\mathbb{D}.X\geq0$ with $X\in C^N$ being a random column vector. In fact, ${\rm Det}(\mathbb{D})\geq0$  turns into the product form uncertainty relation and $X^\dag.\mathbb{D}.X\geq0$ becomes the sum form uncertainty relation. For instance, taking $N=2$ and $X=\{1,\mp i\}^T$, one can obtain that ${\rm Det}(\mathbb{D})\geq0$ is the SUR and $X^\dag.\mathbb{D}.X\geq0$ is the sum form uncertainty relation $\Delta A^{2}+\Delta B^{2}=|\langle [A,B]\rangle|$. Thus, the SUR can be interpreted as the fundamental inequality $\langle \mathcal{F}^\dag \mathcal{F}\rangle\geq0$ or $\mathbb{D}:\geq0$, and so do the other uncertainty relations deduced in Refs. \cite{28LS,52,53,54,55}.

However, the quantum properties of the operator $\mathcal{F}$, in most cases, cannot be fully expressed by $\langle \mathcal{F}^\dag \mathcal{F}\rangle\geq0$, because the non-negativity of the second-order origin moment $\langle \mathcal{F}^\dag \mathcal{F}\rangle\geq0$ cannot provide any information of $\mathcal{F}$ in the quantum level. Considering an arbitrary operator $\mathcal{O}$ , based on the unified uncertainty relation (4), one has:
\begin{align}
\langle \mathcal{F}^\dag \mathcal{F}\rangle\geq\dfrac{|\langle i[{\mathcal{F}},{\mathcal{O}}]_{\mathcal{G}}\rangle|^2+|\langle\{{\mathcal{F}},{\mathcal{O}}\}_{\mathcal{G}}\rangle|^2} {4\langle\mathcal{O}^\dag\mathcal{O}\rangle} \tag{6}.
\end{align}
Especially, we have $|\langle i[{\mathcal{F}},{\mathcal{O}}]_{\mathcal{G}}\rangle|^2+|\langle\{{\mathcal{F}},{\mathcal{O}}\}_{\mathcal{G}}\rangle|^2/4\langle\mathcal{O}^\dag\mathcal{O}\rangle>0$ when the operator $\mathcal{O}$ is generalized-incompatible or generalized-anti-incompatible with $\mathcal{F}$. Obviously, the introduction of $\mathcal{O}$ provides a more accurate description for the second-order origin moment $\langle \mathcal{F}^\dag \mathcal{F}\rangle$. That is to say, the operator $\mathcal{O}$ can provide information for the second-order origin moment of $\mathcal{F}$ that $\langle \mathcal{F}^\dag \mathcal{F}\rangle\geq0$ cannot do, and thus we name  $\mathcal{O}$ as the information operator. In order to investigate the quantum uncertainty relation more accurately, the information operator should be introduced. Using (6), we have:
\begin{align}
&\mathbb{D}:\geq \mathbb{V}\tag{7},
\end{align}
where $\mathbb{V}$ is the $N\times N$ dimension positive semidefinite matrix with the elements $\mathbb{V}(m,n)=\langle\check{A}_m^\dag\mathcal{O}\rangle\langle\mathcal{O}^\dag\check{A}_n\rangle/\langle\mathcal{O}^\dag\mathcal{O}\rangle$ and $\mathbb{D}:\geq \mathbb{V}$ means $\mathbb{D}-\mathbb{V}$ is a positive semidefinite matrix. Based on the properties of the positive semidefinite matrix, we can obtain a series of uncertainty relations for $N$ observables in both the product form and the sum form.

To demonstrate the importance of the information operator, we will investigate its function on fixing the deficiencies appearing in the traditional uncertainty relations. The triviality problem of the  SUR occurs when the state of the system happens to be the eigenstate of $A$ or $B$. For instance, one has $|\langle[A,B]\rangle/2i|^2+|\langle\{\check{A},\check{B}\}\rangle/2|^2\equiv{\Delta A}^2{\Delta B}^2\equiv0$ in the finite-dimension Hilbert space when  ${\Delta A}^2=0$ or ${\Delta B}^2=0$. Different from ${\Delta A}^2{\Delta B}^2$, the sum of the variances ${\Delta A}^2+{\Delta B}^2$ will never equal zero for incompatible observables even when  the state of the system is the eigenstate of $A$ or $B$. Thus the sum form has the advantage in expressing the uncertainty relation. However, the lower bounds of the most sum form uncertainty relations depend on the state $|\psi^\perp\rangle$, making them difficult to apply to the high dimension Hilbert space \cite{21}. Based on the analysis in the previous section, the sum form uncertainty relation (2) can be written as
$\Delta\mathcal{A}^{2}\Delta\mathcal{B}^{2}\geq|\langle\psi|[\mathcal{\check{A}},\mathcal{\check{B}}]_{\mathcal{G}}|\psi\rangle|^{2}/4+|\langle\psi|\{\mathcal{\check{A}},\mathcal{\check{B}}\}_{\mathcal{G}}|\psi\rangle|^{2}/4$, where $\mathcal{A}=A\pm iB$ and $\mathcal{B}=|\psi^\bot\rangle\langle\psi|$, which means the uncertainty relation (2) is still a type of the SUR. Obviously, the state $|\psi\rangle$ will never be the eigenstate of $\mathcal{B}$ when we take $\mathcal{B}=|\psi^\bot\rangle\langle\psi|$, and therefore the triviality problem of the SUR can be remedied by (2) \cite{40}. However, it is due to the existence of $|\psi^\bot\rangle\langle\psi|$ that the uncertainty relation (2) cannot be applied to the high dimension system. Thus, the triviality problem of the product form uncertainty can be considered as the essential reason for the phenomenon that lots of sum form uncertainty relations are difficult to apply to the high dimension system.

In fact, the physical essence of the triviality problem  can be described as that we cannot obtain any information of the uncertainty of $A(B)$ by the product form uncertainty relation, when the state of the system happens to be the eigenstate of $B(A)$. Thus, the information operator, which can provide the information for the uncertainty relation, can be used to fix this triviality problem.  Here, two generalized-incompatible operators $\mathcal{R}$ and $\mathcal{S}$ will be introduced as the information operators. According to (4) and (6), the information operator $\mathcal{O}$ will not contain any effective information of $\mathcal{F}$ when $\langle\mathcal{O}^\dag\mathcal{O}\rangle=0$, and thus the information operator introduced to fix the triviality problem should satisfy $\langle\mathcal{O}^\dag\mathcal{O}\rangle\neq0$. Based on the unified uncertainty relation (4), the second-order origin moments of the generalized-incompatible operators $\mathcal{R}$ and $\mathcal{S}$ will never be zero at the same time, hence at least one of the two information operators can provide effective information to fix the triviality problem. The corresponding uncertainty relation is obtained as (please see the Information operator in the Supplemental Material \cite{35}):
\begin{align}
{\Delta A}^2+{\Delta B}^2\geq&\max_{\mathcal{O}\in\{\mathcal{R},\mathcal{S}\}}\{\frac{|\langle\mathcal{O}^\dag(\check{A}+e^{i\theta}\check{B})\rangle|^2}{\langle\mathcal{O}^\dag\mathcal{O}\rangle}\}-\langle\{\check{A},e^{i\theta}\check{B}\}_{\mathcal{G}}\rangle \tag{8},
\end{align}
where $\theta\in[0,2\pi]$ should be chosen to maximize the lower bound. The triviality problem can be completely fixed by the uncertainty relation (8) for almost any choice of the generalized-incompatible operators $\mathcal{R}$ and $\mathcal{S}$ : choose $\mathcal{R}$ and $\mathcal{S}$  that can avoid $\langle\check{A}\check{B}\rangle\equiv\langle\mathcal{R}^\dag\check{A}\rangle\equiv\langle\mathcal{R}^\dag\check{B}\rangle\equiv\langle\mathcal{S}^\dag\check{A}\rangle\equiv\langle\mathcal{S}^\dag\check{B}\rangle\equiv0$. Such a choice is always possible, as shown in Fig.1.

Due to the absence of  $|\psi^\bot\rangle$, the uncertainty relation (8) can be well applied to the high dimension system. Meanwhile, the uncertainty relation (8) has a tighter lower bound than the uncertainty realtion depending on  $|\psi^\bot\rangle$ by limiting the choice of the information operator, as shown in Fig.1. Furthermore, the inequality (8) will become an equality on the condition that $\mathcal{R}$ or $\mathcal{S}=\lambda_1 \check{A}+\lambda_2 \check{B} $ with ${|\lambda_1|}^2={|\lambda_2|}^2\neq0$ and $\lambda_1,\lambda_2\in C$. The condition is independent on the state $|\psi^\bot\rangle$, and thus can be easily satisfied even for the high dimension Hilbert space. Besides, the uncertainty relation (8) will reduce to the uncertainty relation (2) when taking $\mathcal{R}=|\psi^\bot\rangle\langle\psi|$ and ignoring the influence of the other information operator $\mathcal{S}$, which means that the uncertainty relation (2) can also be considered as taking $|\psi^\bot\rangle\langle\psi|$ as the information operator.

The introduction of the information operator makes us express the uncertainty relation more accurately. Based on the unified uncertainty relation (4), we can obtain the following uncertainty equality (please see the Information operator in the Supplemental Material \cite{35}):
\begin{align}
&\mathbb{D}=\sum^r_{k=1}\mathbb{V}_k \tag{9},
\end{align}
where $\mathbb{V}_k$ is the $N\times N$ dimension positive semidefinite matrix with the elements $\mathbb{V}_k(m,n)=\langle\check{A}_m^\dag\mathcal{O}_k\rangle\langle\mathcal{O}_k^\dag\check{A}_n\rangle/\langle\mathcal{O}_k^\dag\mathcal{O}_k\rangle$,
 $\mathcal{O}_k$ is the element of the operator set $\Theta=\{\mathcal{O}_1,\mathcal{O}_2,\cdots,\mathcal{O}_r\}$ in which the elements satisfy $\langle\mathcal{O}^\dag_i\mathcal{O}_j\rangle=\langle\mathcal{O}^\dag_i\mathcal{O}_j\rangle\delta_{ij}$ and $\langle\mathcal{O}^\dag_k\mathcal{O}_k\rangle\neq 0$  with $k,i,j\in\{1,2,\cdots,r\}$, and $r$ is the maximum number of the elements that the set $\Theta$ can hold. The set can be obtained by the Schmidt transformation (please see the Schmidt transformation process in the Supplemental Material \cite{35})  \cite{LM,NI}. The value of $r$ is equal to the rank of the Metric matrix corresponding to the bilinear operator function$\langle\mathcal{A}^\dagger\mathcal{B}\rangle$, and only depends  on the state of the system. It is worth mentioning that $r$ is less than $d$ for the pure state and less than $d^2$ for the mixed state in the $d$-dimension system, and $r$ will tend to the infinity when considering the infinite-dimension system. The uncertainty equality indicates that the information of the uncertainties for incompatible observables can be captured accurately when $r$ information operators are introduced, as shown in Fig.2.

\emph{Discussion.--- }The variance-based uncertainty relations can be divided into the product form and the sum form. The product form uncertainty relation cannot fully capture the concept of the incompatible observables, and the problem is referred to as the triviality problem of the product form uncertainty relation. The triviality problem can be fixed by the sum form uncertainty relation, and thus lots of effort has been made to investigate the sum form uncertainty relation. However, most of the sum form uncertainty relations depend on the orthogonal state to the state of the system, and are difficult to apply to the high dimension Hilbert space.

We provide a unified uncertainty relation for the two forms uncertainty relations, and deduce that the essences of the product form and the sum form uncertainty relations are actually the unified uncertainty relation for Hermitian operators and non-Hermitian operators, respectively.
Thus, the unified uncertainty relation provides a unified framework for the two forms uncertainty relations.

In the unified framework, we deduce that the uncertainty relation for incompatible observables is bounded  by not only the commutator of themselves, but also the quantities related with the other operator, which can provide information for the uncertainty and thus is named as the information operator. The deficiencies in the product form and the sum form uncertainty relations are actually identical in essence, and can be completely fixed by the introduction of the information operators. Furthermore, the uncertainty inequality will become an uncertainty equality when a specific number of information operators are introduced, which means the uncertainty relation can be expressed exactly with the help of the information operators. Thus, the unified framework provides a strengthened theoretical system for the uncertainty relation.

The unified framework also provides a new interpretation of the quantum uncertainty relation for the non-Hermitian operators, i.e., the ``observable" second-order origin moments of the non-Hermitian operators cannot be arbitrarily small at the same time when they are generalized-incompatible or generalized-anti-incompatible with each other. The new interpretation reveals some novel quantum properties that the traditional uncertainty relation cannot do

This work is supported by the National Natural Science Foundation of China (Grant Nos.11574022, 61227902, 11774406, 11434015, 61835013), MOST of China (Nos. 2016YFA0302104, 2016YFA0300600), the Chinese Academy of Sciences Central of Excellence in Topological Quantum Computation (XDPB-0803), the National Key R\&D Program of China under grants Nos. 2016YFA0301500, SPRPCAS under grants No. XDB01020300, XDB21030300.

X. Z. and S. Q. M. contributed equally to this work.


\begin{thebibliography}{99}
\bibitem{25F}To make sure that the quantity measuring the uncertainty will be a real number, the variance is taken as $\langle(Q-\langle Q\rangle)^\dag(Q-\langle Q\rangle)$ for non-Hermitian operators. Here the $\langle Q\rangle$ represents the expected value of $Q$.
\bibitem{40}The uncertainty turns into $\Delta A^2+\Delta B^2=\pm i\langle[A,B]\rangle$ when the state of system is chosen as the eigenstates of $\mathcal{A}=A\pm iB$, which is guranteed to be nontrivial.
\end{thebibliography}
\end{document}